\documentstyle[12pt,epsfig]{article}

\begin{document}
\newcounter{num}
\def\lsim{\mathrel{\lower4pt\hbox{$\sim$}}\hskip-12pt\raise1.6pt\hbox{$<$}\;}
\def\BAR{\bar}
\def\fm{{\cal M}}
\def\fl{{\cal L}}
\def\fp{{\cal P}}
\def\n{{n}}
\def\delt{{n}}
\def\gsim{\mathrel{\lower4pt\hbox{$\sim$}}
\hskip-10pt\raise1.6pt\hbox{$>$}\;}

\vspace*{-.5in}
\rightline{AMES-HET 99-06}   
\rightline{UCRHEP-T258}
\rightline{BNL-HET-99/13}
\begin{center}

{\large\bf 
Gauge Boson -- Gauge Boson Scattering in Theories with Large Extra Dimensions}

\vspace{.3in}

David Atwood$^{1}$\\ 
\noindent Department of Physics and Astronomy, Iowa State University, Ames,
IA\ \ \hspace*{6pt}50011\\
\medskip

Shaouly Bar-Shalom$^{2}$\\ 
\noindent Department of Physics, University of California, Riverside, 
CA 92521\\
\medskip

\medskip
and\\
\medskip

Amarjit Soni$^{3}$\\
\noindent Theory Group, Brookhaven National Laboratory, Upton, NY\ \ 
11973\\

\footnotetext[1]{email: atwood@iastate.edu}
\footnotetext[2]{email: shaouly@phyun0.ucr.edu}
\footnotetext[3]{email: soni@bnl.gov}
\end{center}
\vspace{.2in}

\begin{quote}
{\bf Abstract:}
  
We consider the scattering amplitudes of the form 
$V_1V_2\to V_3V_4$, where $V_i=\gamma,~Z,~W$ or $g$(=gluon) are the 
Standard Model gauge bosons, 
due to graviton exchange in Kaluza-Klein theories with large extra dimensions.
This leads to a number of experimentally viable signatures at high energy
leptonic and hadronic colliders. 
We discuss the observability or  
attainable limits on the scale of the gravitational interactions ($m_D$),
that may be obtained at an $e^+e^-$ Next Linear Collider (NLC) and at the 
LHC, by 
studying some of these type of gauge boson scattering processes.   
We find that the attainable limits through these type of processes are: 
$m_D \gsim 3$ TeV at the NLC and $m_D \gsim 6$ TeV at the LHC.

\end{quote} 
\vspace{.3in} 
\newpage

%
%
%
%
%
%

\section{Introduction}

Although gravity is the weakest force in nature and naive dimensional
arguments suggest that its effect on high energy collisions is
insignificant below energies of the Planck scale
$M_P=G_N^{-1/2}=1.22\times 10^{19}~GeV$, recent advances in
M-theory~\cite{mtheory} have motivated the consideration of
Kaluza-Klein theories which allow gravitational interactions to become
strong at relatively modest energies compared to the Planck scale, perhaps
as low as $1~TeV$~\cite{add1,add2}.

M-theory is a special case of Kaluza-Klein theories where there are total
of 11 dimensions. Four of these are of course the usual space time while
the other 7 are compact.  The possibility which emerges
from~\cite{add1,add2} is that while some of the compact dimensions have
lengths near the Planck scale, $\delt$ of these dimensions may be
compactified with a distance scale $R$ which is much larger.  This leads
to an effective Planck mass $m_D$, perhaps of the order of $1~TeV$ which
is related to the size $R$ of the new dimension according
to~\cite{add1}: 

\begin{eqnarray}
8\pi R^\delt m_D^{2+\delt}\sim M_P^2 ~.
\label{Rsize}
\end{eqnarray}

\noindent 
In this scenario, at distances $d<R$ the Newtonian inverse square law will
fail~\cite{add1}. If $\delt=1$ and $m_D$=$1~TeV$, then R is of the order
of $10^{8}~km$, large on the scale of the solar system, which is clearly
ruled out by astronomical observations.  However, if $\delt\geq 2$ then
$R< 1~mm$; there are no experimental constraints on the behavior of
gravitation at such scales~\cite{cavin} so these models 
may not be inconsistent with experimental results.

Astonishingly enough if $m_D\sim 1~TeV$ then gravitons may be readily
produced in accelerator experiments.  This is because the extra dimensions
give an increased phase space for graviton radiation. Another way of
looking at this situation is to interpret gravitons which move parallel to
the 4 dimensions of space time as the usual gravitons giving rise to
Newtonian gravity while the gravitons with momentum components
perpendicular to the brane are effectively a continuum of massive objects.
The density of gravitons states is given by~\cite{add1,add2,wells,taohan}:

\begin{eqnarray}
D(m^2)={dN\over d m^2}={1\over 2} S_{\delt-1} 
{\BAR M_P^2 m^{\delt-2}\over m_D^{\delt+2}} ~,
\end{eqnarray}

\noindent
where $m$ is the mass of the graviton, $\BAR M_P=M_P/\sqrt{8\pi}$
and 
$S_k=2\pi^{(k+1)/2}/\Gamma[(k+1)/2]$.
The probability of graviton emission  may thus become large when the sum
over the huge number of graviton modes is considered. 
Of course this distribution cannot increase in this way forever. At 
energies $> m_D$ the effects of the fundamental theory should become 
manifest and so we will suppose that the distribution is cutoff at $\sim 
m_D$.

Gravitons with polarizations that lie entirely within the physical
dimensions are effective spin 2 objects.  Gravitons with polarizations
partially or completely perpendicular to the physical brane are vector and
scalar objects. Processes that are sensitive to the scalar states are of
particular interest because the scalar couplings are proportional to mass
and so are often weakly coupled to processes exclusively involving
particles of low mass.

The prospect that a realistic compactification of M-theory leads to
processes that can be readily observed has lead to considerable
phenomenological activity~\cite{wells}-\cite{gammagamma_ZZ}. 
Broadly speaking there are two
kinds of processes where the effects due to this form of gravitation may
be detected. First of all, a real graviton may be produced which leaves
the detector resulting in a signature involving missing mass and energy, see e.g., \cite{real,ABS1}.
One advantage of this class of reactions is that if a signal is seen, the
missing mass spectrum would give strong evidence that these gravity
theories are involved and indicate the value of $\delt$. The disadvantage
is that the rates tend to be suppressed at large $\delt$ due to the lower
density of states; hence the limits that may be set on $m_D$ tend to be
less restrictive at larger $\delt$. 

Secondly, there are processes mediated by virtual gravitons, see e.g., 
\cite{virtual,4gammapaper,gammagamma_ZZ}.  When a
virtual graviton is exchanged, each of the graviton states adds to the
amplitude coherently, thus in the sum over gravitons the density of states
cancels the $1/M_P^2$ from the gravitational coupling.  The disadvantage
of these processes is that if a signal is seen, it is unlikely to be easy to
prove that gravitational interaction is responsible as opposed to some
other new physics. Of course if a limit is being set by the absence of a
signal, this is not a problem.  The advantage of these processes is that
the whole tower of graviton states acts coherently so the results are
largely independent of $\delt$. Limits can thus be set on all values of
$\delt$ simultaneously. In this paper we will consider process which are
of the latter type.

In any virtual process, there will in general exist some Standard Model (SM)
background. Clearly it is beneficial to choose processes where the
SM background is so small that it does not limit the bound that
can be placed on $m_D$. The class of process which we consider here are of
the form $V_1V_2\to V_3 V_4$ where $V_i$ 
are SM gauge bosons which may be distinct or identical.
If the tree level coupling between these four specific bosons does not
follow from the gauge theory, then the scattering proceeds only at fourth
order in the gauge coupling and is therefore highly suppressed as is the 
case e.g. for $\gamma\gamma\to \gamma\gamma$, $ZZ\to ZZ$ or 
$gg \to \gamma \gamma$, $gg \to ZZ$.

%
%
%

For each of the virtual processes
of the form $V_1V_2\to V_3V_4$ which involve the exchange of gravitons, 
we need to sum the propagator over all 
of the possible graviton states. 
The amplitudes 
considered here can be 
factored into one of the following forms:

\begin{itemize}
\item[(a)]
$\fm=\left ( i/( s-m^2) \right )\kappa^2 \hat \fm_s~,$
\item[(b)]
$\fm=\left ( i/ (t-m^2) \right )\kappa^2 \hat \fm_t~,$
\item[(c)]
$\fm=\left ( i/ (u-m^2) \right )\kappa^2 \hat \fm_u~,$
\end{itemize}

\noindent
where 
$m$ is the mass of the graviton and 
all the $m$ dependence is in the propagator factor. 
Also, $\kappa=\sqrt{16 \pi G_N}$ \cite{taohan}, where $G_N$ is 
the Newtonian gravitational constant.  

In case (a), for example, the total amplitude including all graviton 
exchanges is thus:

\begin{eqnarray}
\fm^{tot}_s=\hat\fm_s  \sum_\nu {i\over s-m_\nu^2}~,
\end{eqnarray}

\noindent 
where $\nu$ indexes the graviton masses $m_\nu$. 
We write the sum:

\begin{eqnarray}
\sum_\nu {i\over s-m_\nu^2}=D(s) ~,
\end{eqnarray}

\noindent
where the value of $D(s)$ calculated in~\cite{wells,taohan}
is:

\begin{eqnarray}
\kappa^2 D(s)=
-i{16\pi\over m_D^4}F + O({s\over m_D^2}) ~.
\end{eqnarray}

\noindent 
The constant $F$ contains all the dependence on $n$ and is given by:

\begin{eqnarray}
F=\left \{
\begin{array}{cl}
\log(s/m_D^2)    & ~~~{\rm for}~n=2\\
2/(n-2)          & ~~~{\rm for}~n>2
\end{array}
\right .
\label{eqf}~.
\end{eqnarray}

\noindent Likewise for the $t$-channel, we can define: 

\begin{eqnarray}
\sum_\nu {i\over t-m_\nu^2}=D_E(t) ~,
\end{eqnarray}

\noindent
and similarly for the $u$-channel.
In the case of a $2\to 2$ process, to lowest order in $s/m_D^2$,  
$D_E(t)=D_E(u)=D(s)$ \cite{taohan}. 
Thus, in general, the sum of all three channels is: 

\begin{eqnarray}
\fm \approx \kappa^2 D(s) 
(
\hat\fm_s+
\hat\fm_t+
\hat\fm_u
)
\approx
-i{16\pi\over m_D^4}F 
(
\hat\fm_s+
\hat\fm_t+
\hat\fm_u
) ~.
\end{eqnarray}

\noindent 
Defining $z=\cos\theta$ where $\theta$ is the angle between $V_1$ and 
$V_3$ in the cms frame, the differential cross section is thus given by: 

\begin{eqnarray}
{d\sigma\over d z}
=
\left (
{8\pi F^2\over s m_D^8 B \fp}  
\right )
\left (
{2 | \vec P_3 | \over\sqrt{s}} 
\right )
\sum_{polarization,color}
\left |
\hat\fm_s+
\hat\fm_t+
\hat\fm_u
\right |^2
\end{eqnarray}

\noindent
where $B=2$ for identical final state particles and  1 otherwise while $\fp$ 
is the number of initial color times polarization states averaged over.

\section{Gauge Boson Scattering}

Let us first enumerate some of the instances of this kind of scattering 
which can be of interest. We will break it down into the following 
categories: \begin{itemize}
\item
$Z$ and $\gamma$ only:
\begin{eqnarray}
\begin{array}{ll}
{\rm (a)}~~~\gamma\gamma\to\gamma\gamma
~~~~~~&~~~~~~
{\rm (b)}~~~\gamma\gamma\to ZZ
\\
{\rm (c)}~~~ZZ\to\gamma\gamma 
~~~~~~&~~~~~~
{\rm (d)}~~~\gamma Z\to\gamma Z
\end{array}
\end{eqnarray}

\item
2 $W$'s with $Z$, $W$ and $\gamma$:
\begin{eqnarray}
\begin{array}{ll}
{\rm (e)}~~~\gamma\gamma\to W^+W^- ;~ W^+W^-\to\gamma\gamma
&
{\rm (f)}~~~W\gamma\to W\gamma
\\
{\rm (g)}~~~ZZ\to W^+W^- ;~ W^+W^-\to ZZ
&
{\rm (h)}~~~WZ\to WZ
\\
{\rm (i)}~~~W^+W^-\to W^+W^- ;~ W^-W^-\to W^-W^-
&
\ 
\end{array}
\end{eqnarray}

\item
Processes with 2 gluons
\begin{eqnarray}
\begin{array}{ll}
{\rm (j)}~~~gg\to\gamma\gamma;~\gamma\gamma\to gg
~~~~~~&~~~~~~
{\rm (k)}~~~g\gamma\to g\gamma
\\
{\rm (l)}~~~gg\to ZZ;~ZZ \to gg
~~~~~~&~~~~~~
{\rm (m)}~~~gZ \to gZ 
\\
{\rm (n)}~~~gg\to WW;~WW \to gg
~~~~~~&~~~~~~
{\rm (o)}~~~gW \to gW 
\end{array}
\end{eqnarray}

\item
Four gluon coupling:
\begin{eqnarray}
\begin{array}{ll}
{\rm (p)}~~~gg\to gg
~~~~~~&~~~~~~
\ 
\end{array}
\end{eqnarray}

\item
Four $Z$ coupling:
\begin{eqnarray}
\begin{array}{ll}
{\rm (q)}~~~ZZ\to ZZ
~~~~~~&~~~~~~
\ 
\end{array}
\end{eqnarray}

\end{itemize}

These processes will proceed through the Feynman
diagrams shown in Fig.~1 where the dashed line represents a spin 2 or 
spin 0 graviton. The exchange may be in various combinations of the 
$s$, $t$ and $u$-channels  
depending on what the external bosons are. The spin 0 exchange is only 
operative if all of the bosons are massive, specifically (g), (i), (h), 
and (q). 

Diagrams which contain either 4 gluons as in case (p),  
at least 2 $W$'s as in cases (e)-(h) and 4 $Z$'s as in case (q) can proceed 
through tree level SM processes and so the possible large 
backgrounds must be considered.
The other processes will not have tree level SM backgrounds. 
In what follows we will not consider any process involving 
$W$-bosons such as in cases (e)-(h), (n) and (o). We will also not 
explicitly consider the processes $ZZ \to gg$ (case (l)) and 
$gZ \to gZ$ (case (m)). We note, however, 
that the formulae we give in the appendices can be easily generalized to
include those processes for future use.  

\begin{figure}[htb]
\psfull
 \begin{center}
 \leavevmode
 \epsfig{file=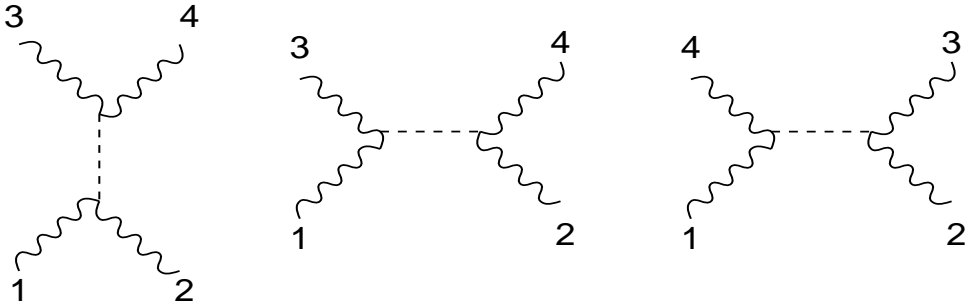,height=6cm,width=6cm,bbllx=0cm,bblly=2cm,bburx=20cm,bbury=25cm,angle=0}
 \end{center}
\caption{\emph{The three Feynman diagrams that give rise 
to gauge boson scattering through virtual graviton exchange. 
Dashed lines stand for the exchanges of a spin 2 or a spin 0 
graviton.}}
\label{fig1}
\end{figure}



\section{Electron-Positron Colliders}  

To experimentally study the scattering of gauge bosons it is usually 
necessary to 
consider reactions where the initial state consists of fermions, in 
particular 
$e^+$, $e^-$ $p$ or $\bar p$. At high energies   
virtual gauge bosons are then produced nearly on shell and collinear with 
the initial particles. If one uses the effective boson 
approximation~\cite{evba,collider} 
one may regard the fermion beams as sources of 
gauge bosons and at leading log ignore the virtuality of these bosons in 
calculating the cross section. 

At electron-positron colliders a number of reactions of 
this type involving bosons may be studied. The simplest is perhaps 
$\gamma\gamma\to\gamma\gamma$. 
It should be noted, however, that 
an electron-positron collider can be converted into 
an almost monochromatic photon-photon 
collider using back-scatter laser technique \cite{photoncollider}.
An NLC running in such a photon mode has an advantage 
that the initial photons may be polarized. 
In fact, the reaction $\gamma\gamma\to\gamma\gamma$ 
has been considered before in~\cite{4gammapaper} 
in the context of a photon-photon collider. There it was shown that 
the discovery reach of low energy gravity through the 
process $\gamma\gamma\to\gamma\gamma$ may be considerably improved 
if the initial photons are polarized. The same sensitivity to the initial 
photons polarization was also found in \cite{gammagamma_ZZ} for the reactions
$\gamma \gamma \to W^+W^-,~ZZ$. 
 
Here, we wish instead to explore yet another aspect of this type of 
$V_1 V_2 \to V_3 V_4$ scattering processes, 
namely, virtual gauge boson emission 
from the initial $e^+e^-$ of a NLC running in its ``simple'' 
mode\footnote{For definiteness we discuss 
gauge boson scattering in an electron-positron collider. Our results below, 
however, are clearly extendable 
to muon colliders as well.}.
Clearly, in the case of $V_1=V_2=\gamma$, 
$\gamma\gamma\to VV$ will then lead to a
signature of the form $e^+ e^- \to  VV e^+ e^-$ which is different 
from the hard process $\gamma\gamma\to VV$ (directly observable in 
a photon-photon collider) and has its own kinematic characteristics. 

For instance, in the case of an electron-positron 
collider, two photon 
processes are thus calculated using the equivalent photon approximation so 
that if a cross section $\sigma_{\gamma\gamma\to X}$ is known, the cross 
section for $e^+e^-\to e^+e^-+X$ via the two photon mechanism is given by:

\begin{eqnarray}
\sigma_{e^+e^-\to e^+e^-X}
=
\left (
{\alpha\over 2\pi}
\log (s_0/4 \hat m_e^2)
\right )^2
\int_0^1 f(\tau) \sigma_{\gamma\gamma\to X}(\tau s_0)~d\tau ~,
\end{eqnarray}

\noindent 
where $s_0$ is the square of the center of mass energy of the initial 
$e^+e^-$ and:

\begin{eqnarray}
f(\tau)={1\over \tau}((2+\tau)^2\log{1\over \tau} - 2(1-\tau)(3+\tau)) ~.
\end{eqnarray}

\noindent 
In this expression the total cross section for $e^+e^-\to e^+e^-+X$ is 
given if one takes $\hat m_e=m_e$ as the mass of the electron. If one 
wishes, however to observe the $e^+e^-$ in the final state, experimental 
considerations suggest that a minimum cut on the transverse momentum of 
the final state electrons $P_{Tmin}$ be used. In this case the result is 
given by taking $\hat m_e=P_{Tmin}$.
For instance, if one is considering the production of real gravitons as 
in~\cite{ABS1}, it is essential to observe these final state electrons 
since the graviton itself is undetectable. 

\begin{figure}[htb]
\psfull
 \begin{center}
 \leavevmode
 \epsfig{file=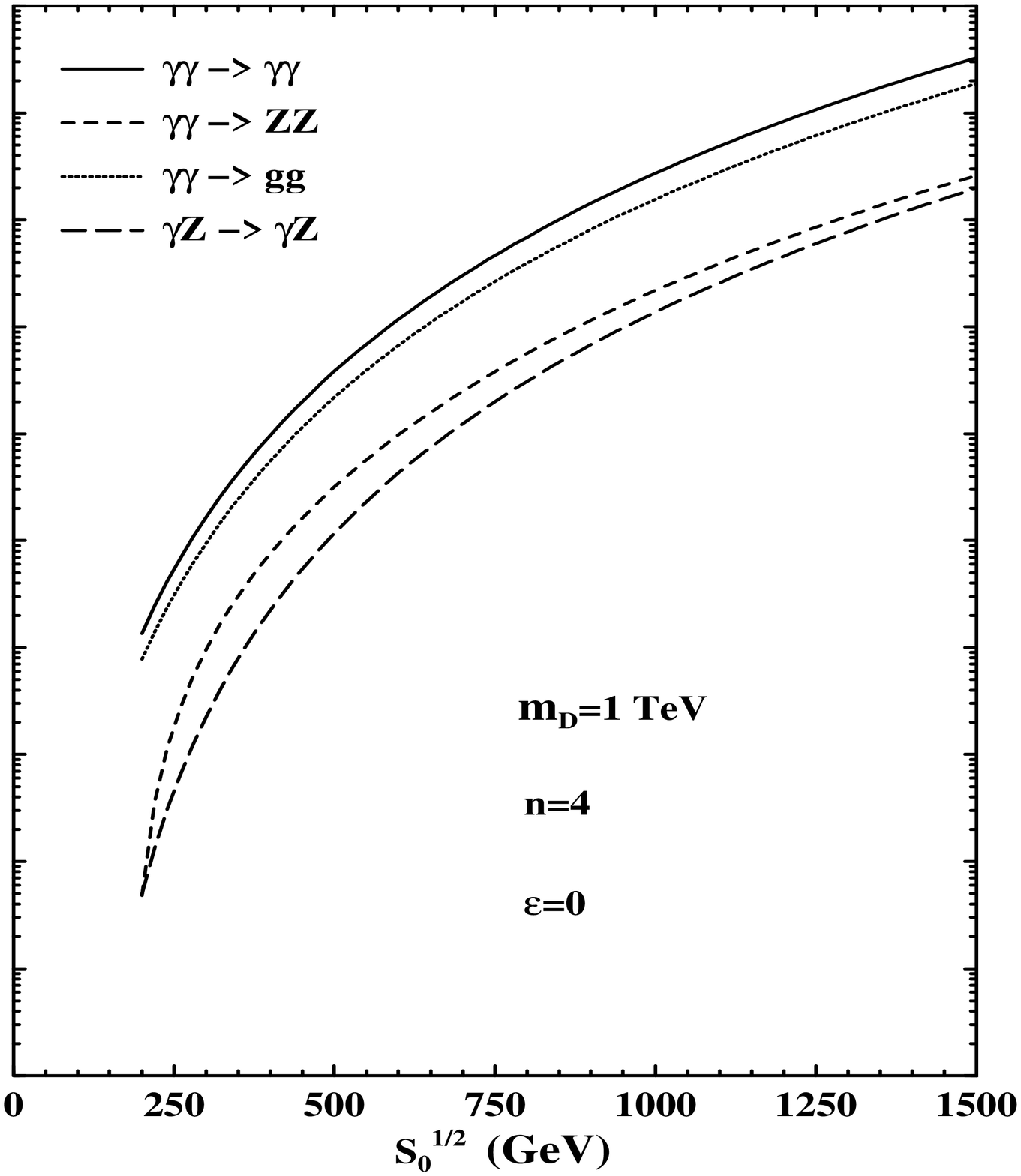,height=8cm,width=8cm,bbllx=0cm,bblly=2cm,bburx=20cm,bbury=25cm,angle=0}
 \end{center}
\caption{\emph{The overall cross sections $\sigma(e^+e^- \to V_1 V_2 e^+e^-)$, 
where $V_i=\gamma,~Z$ or $g$, for various channels of effective gauge boson
scattering sub-processes in an electron-positron collider, as a function of
 the center of mass energy of the collision, $\sqrt{s_0}$.
In all the curves we take $m_D=1~TeV$ and $n=4$.
The $\gamma\gamma\to\gamma\gamma$ sub-process is shown with a solid line,
the $\gamma\gamma\to ZZ$ sub-process is shown with a short dashed line,
the $\gamma\gamma\to gg$ sub-process is shown with a dotted line and
the $\gamma Z\to \gamma Z$ sub-process is shown with a long dashed line.}}
\label{fig2}
\end{figure}

It is well known~\cite{evba,collider} 
that at high energy colliders this expression can be generalized to 
cases where the photons are replaced with $W$ or $Z$ bosons; in 
those cases one must generally use helicity dependent structure functions 
for the initial state gauge bosons. 

\begin{figure}[htb]
\psfull
 \begin{center}
 \leavevmode
 \epsfig{file=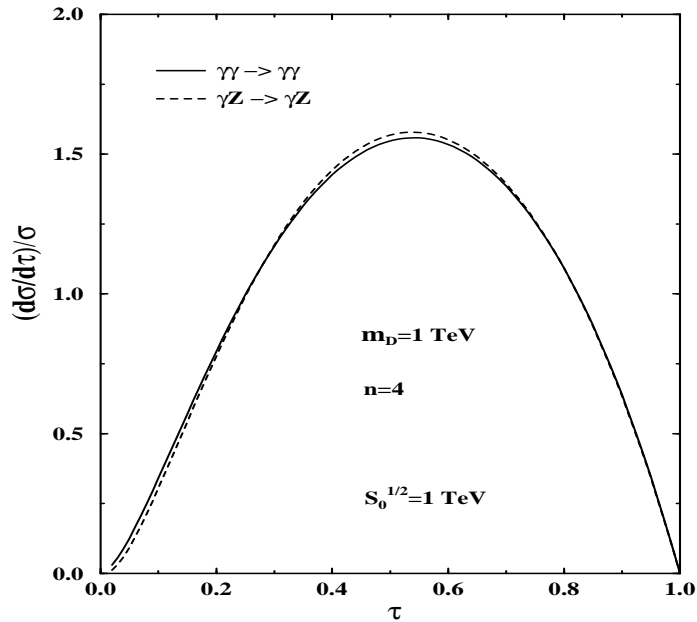,height=8cm,width=8cm,bbllx=0cm,bblly=2cm,bburx=20cm,bbury=25cm,angle=0}
 \end{center}
\caption{\emph{The normalized invariant mass 
distribution $(d\sigma/d\tau)/\sigma$ is shown 
for $e^+e^- \to \gamma \gamma e^+e^-$ via the sub-process 
$\gamma\gamma\to\gamma\gamma$ (solid 
line) and for $e^+e^- \to \gamma Z e^+e^-$ via the sub-process
$\gamma Z\to \gamma Z$ (dashed line), where $\sqrt{s_0}=1~TeV$. The other 
model parameters are the same as in Fig.~\ref{fig2}.}}
\label{fig3}
\end{figure}

The helicity amplitudes and cross section
for $\gamma \gamma \to \gamma \gamma$ are given in appendix A 
(eqs.~(\ref{a1})--(\ref{a3}))   
and 
the corresponding full 
cross section, $\sigma(e^+e^- \to \gamma \gamma e^+e^-)$, is 
shown in Fig.~\ref{fig2} as a function of $\sqrt s_0$ for the case where 
$n=4$ (or equivalently $F=1$, see 
eq.~(\ref{eqf})) and $m_D=1~TeV$.

In Fig.~\ref{fig3} 
we show the normalized distribution of $(d\sigma/d\tau)/\sigma$, 
where $\tau=s/s_0$,
from which it is clear that most of these events are concentrated at
high invariant $\gamma \gamma$ mass.  
Thus a large portion of the events that make up this cross section
would be quite distinctive.  This follows from the fact that the cross
section is proportional to $s^3$; hence, even though the effective
luminosity distribution of the photons decreases at large $\tau$, the
growth of the cross section with $\tau$ makes the average 
value of $\tau$ become large. The same trend is
also true for all of the other reactions considered and so experimental
tests for these type of processes
 should focus on final states of large invariant
mass.

Since $\sigma(\gamma \gamma \to \gamma \gamma)$ (and therefore 
$\sigma(e^+e^- \to \gamma \gamma e^+e^-)$) 
is proportional to $F^2/m_D^8$, the value of the 
cross section may be trivially adjusted for other values of these input 
parameters; the same is also true for all the other processes which we 
consider below. 
Thus, for instance, if $\sqrt{s_0}=1~TeV$, then the cross section 
is $\sim 270~fb$ with $m_D=1$ TeV, 
so at such an accelerator with a luminosity of $200~fb^{-1}$, 
the limit which could be placed on $m_D$ with a criterion of 10 events (i.e., 
assuming that such a signal is observable only if 10 or more events 
are seen)  
would be $m_D\sim 2.9~TeV$. 
This limit is weaker than the one achievable 
in a photon-photon collider using polarized initial photons by 
about a factor of two (see \cite{4gammapaper}), but is independently useful 
for an electron-positron collider.

The amplitude for $\gamma\gamma\to gg$ is simply the $s$-channel of the
$\gamma\gamma\to\gamma\gamma$ graph. 
The helicity amplitudes and cross section for
this process are given in appendix B 
(eqs.~(\ref{b1})--(\ref{b3})). 
The cross section $\sigma(e^+e^- \to gg e^+e^-)$ 
is also shown in Fig.~\ref{fig2}; it is similar to the
$\gamma\gamma$ final state and in principle leads to similar sensitivity
to $m_D$. The events in this case would consist of two jets which would
typically have a high invariant mass as with the $\gamma\gamma$ events
above. 

Another class of reactions which may be of interest at electron-positron 
colliders is reactions which contain $Z$-bosons in the initial and/or final 
states. For instance $\gamma\gamma\to ZZ$ which has been studied 
in~\cite{gammagamma_ZZ} in the context of a photon--photon collider.
We note again that the NLC in its photon mode 
and with polarized initial photons can place stringer limits on 
$m_D$ \cite{gammagamma_ZZ} by studying $\gamma\gamma\to ZZ$. 
Nonetheless, as shown below, it is clearly important to  
analyze this process in an electron-positron collider as well. 
The helicity amplitudes and cross section for 
this process are 
given in appendix D
(eqs.~(\ref{d1})--(\ref{d3})) and the cross section is plotted
in Fig.~\ref{fig2}. 
This process is smaller than $\gamma\gamma\to \gamma\gamma$ since it only 
proceeds through one channel and also smaller than the similar 
$\gamma\gamma\to gg$ by roughly the color factor of $8$. It does, however, 
have the prospect 
of providing
information  about the polarization of the $Z$'s from 
analysis of their decay distributions. 
In the limit of large $s>>m_Z^2$, the ratio of longitudinal Z-pairs in 
comparison to transverse pairs approaches $1/12$ if indeed gravitational 
interactions are responsible for $\gamma \gamma \to ZZ$. In particular:

\begin{eqnarray}
\left ( {Z_L Z_L\over Z_T Z_T}\right )_{graviton}
=
{1\over 12}(1+4x_Z)^2 \label{ratio}~,
\end{eqnarray}

\noindent
where $x_Z=m_Z^2/s$. 
This may be useful in distinguishing Z pairs produced by gravitons from 
those produced by other new physics mechanisms. For instance, Higgs or 
other massive scalar can be produced in the $s$-channel via 
$\gamma \gamma$ fusion and  
decay to a $ZZ$ pair. This  
would produce a predominance of longitudinally polarized $Z$'s
given by~\cite{higgs_hunters}:\footnote{As discussed below, there is, however, 
a tree level SM background to the $ZZe^+e^-$ final state coming from 
exchanges of the SM Higgs boson in the $ZZ\to ZZ$ sub-process. Such 
Higgs exchanges in the $t$ and $u$-channel will have a different 
(from the one given in eq.~(\ref{ratio})) ratio 
of $Z_LZ_L/Z_TZ_T$. However, since the $ZZ$ luminosity functions are much 
smaller than the $\gamma \gamma$ ones, we expect 
the ratio $Z_LZ_L/Z_TZ_T$ to be dominated by the 
gravity mediated $\gamma \gamma \to ZZ$ sub-process if indeed the 
gravity scale is at the 1 TeV level. Moreover, as further mentioned below, 
the processes $\gamma \gamma \to ZZ$ and $ZZ \to ZZ$
may be distinguished by studying the final state electrons.}

\begin{eqnarray}
\left ( {Z_L Z_L\over Z_T Z_T}\right )_{scalar}
=
{(2-x_Z)^2\over 2 x_Z^2} ~.
\end{eqnarray}

\noindent One could also consider the reverse process 
$ZZ\to \gamma\gamma$; however,  
this would probably not be of experimental interest since the smaller 
$ZZ$ luminosity function would cause it to be about 100 times smaller 
than the $\gamma\gamma\to \gamma\gamma$ process with the same final 
state. 
The crossed graph $\gamma Z\to \gamma Z$, however, could have a significant 
cross section similar to the case of $\gamma\gamma\to ZZ$.\footnote{We 
note that, for massive vector bosons,
 the effective vector boson approximation in leading log 
tends to over estimate the cross section, in particular, 
the cross section coming from fusion of transversely polarized gauge bosons, 
see e.g., Johnson {\it et al.} in \cite{evba}. 
Therefore, the actual overall cross section from $\gamma Z \to \gamma Z$ may 
be slightly smaller than what is shown in Fig.~\ref{fig2}. 
For the point we are making, however, the leading log approximation suffices.}
In this case 
the electron which emits the $Z$ would typically have transverse 
momentum $P_T\sim m_Z$ and hence would be easily detectable providing 
additional constraints on the kinematics of the final state.

Since both $\gamma\gamma\to ZZ$ and $\gamma Z\to \gamma Z$ are about 10
times smaller than $\gamma\gamma\to \gamma\gamma$ these modes would not be
primarily useful in putting a bound on $m_D$. However, if a graviton signal
were seen in the $\gamma\gamma\to\gamma\gamma$ channel, the ratio
$\gamma\gamma:\gamma Z:ZZ$ would be useful in indicating that
gravitational interactions were indeed the explanation since this ratio
would be independent of $m_D$ and $n$ in the propagator summation
approximation discussed above.

The process $ZZ\to ZZ$ potentially has the unique feature that the
exchanged graviton can be either scalar or tensor.  We find, however,
that numerically the cross section in the context of an $e^+e^-$ collider
is not greatly sensitive to the scalar sector.  
This is partly because there is a similar contribution 
to the scalar exchanges in Fig.~\ref{fig1} coming from 
the SM Higgs boson.
The helicity amplitudes for the process $ZZ \to ZZ$, for both the 
graviton and the SM Higgs exchanges are given in appendix G.  

In fact, even assuming for simplicity 
an infinitely heavy Higgs, i.e., disregarding 
the SM background to this process, the 
scalar graviton contribution is rather small as compared to the spin 2 
graviton. In particular,  
in order to test the
sensitivity to the scalar sector in $ZZ\to ZZ$ in the absence of the SM 
diagrams,  
we can multiply the scalar propagator
by $(1+\epsilon)$, where $\epsilon$ is an arbitrary constant. 
Using this, the helicity amplitudes for this 
process, due to the spin 0 (and the spin 2) graviton exchanges are given in
appendix G. 
Note that the only term proportional to $R$ (hence sensitive to the 
scalar graviton), 
which is not suppressed at large $s$ by a factor of $x_Z=m_Z^2/s$, is that 
corresponding to the $0000$ helicity combination. This is because the 
coupling of the scalar graviton to a $ZZ$ pair 
is explicitly 
proportional to the mass-squared of the $Z$ ~\cite{taohan}.
However, when this is coupled to a 
longitudinal state, this dependence is canceled by the 
explicit $1/m_Z$ 
mass dependence due to each of 
the four longitudinal polarization vectors.

To get an idea of the sensitivity of the graviton cross section to $\epsilon$  
let us define: 

\begin{eqnarray}
r(\epsilon)={\sigma(\epsilon)\over \sigma(\epsilon=0)}-1 ~.
\end{eqnarray}

\noindent
Thus, if $n=4$, then for $s_0=1~TeV$, $r(1)=0.009$, $r(5)=0.064$, and
$r(10)=0.18$. 
On the other hand,
if $\sqrt{s_0}=500~GeV$, then $r(1)=0.05$, $r(5)=0.31$ and
$r(10)=0.80$.  Therefore, probably a non-zero value of $\epsilon$ leads to
unobservably small effects unless $\epsilon$ is fairly large, $\epsilon
>>1$. The reason for this is that the contribution from the scalar
exchange is dominated by the case where the initial bosons are
longitudinal but the kernel for $e\to eZ_L$ does not receive logarithmic
enhancement as in the case for transverse $Z$ emission. The scalar
graviton contribution to
the final cross section in the overall reaction is thus modest even though
the hard cross section for $Z_LZ_L\to Z_LZ_L$ via scalar exchange is comparable
to the hard cross section for $Z_TZ_T\to Z_TZ_T$ via tensor exchange.
Likewise, if one measures the proportion of longitudinal and transverse Z 
bosons in the final state, the sensitivity to the scalar exchanged may be 
increased, however, in reality 
the SM Higgs contribution will probably dominate the scalar dynamics 
in this process.

The final state in the case of $ZZ\to ZZ$ is the same as that of
$\gamma\gamma\to ZZ$; however, the two processes may be separated by
observation of the final state electrons. For instance, 
if we impose the
cut that $P_T>m_Z$ for each of the final state electrons at
$\sqrt{s_0}=1~TeV$, the $\gamma\gamma\to ZZ$ is reduced by a factor of
about 70 while the signal due to $ZZ\to ZZ$ 
is reduced by a factor of about 1.3. Using this
cut then, the contribution of $ZZ\to ZZ$ may thus be enhanced relative 
to $\gamma\gamma\to ZZ$. 

\section{Hadronic Colliders}

\begin{figure}[htb]
\psfull
 \begin{center}
 \leavevmode
 \epsfig{file=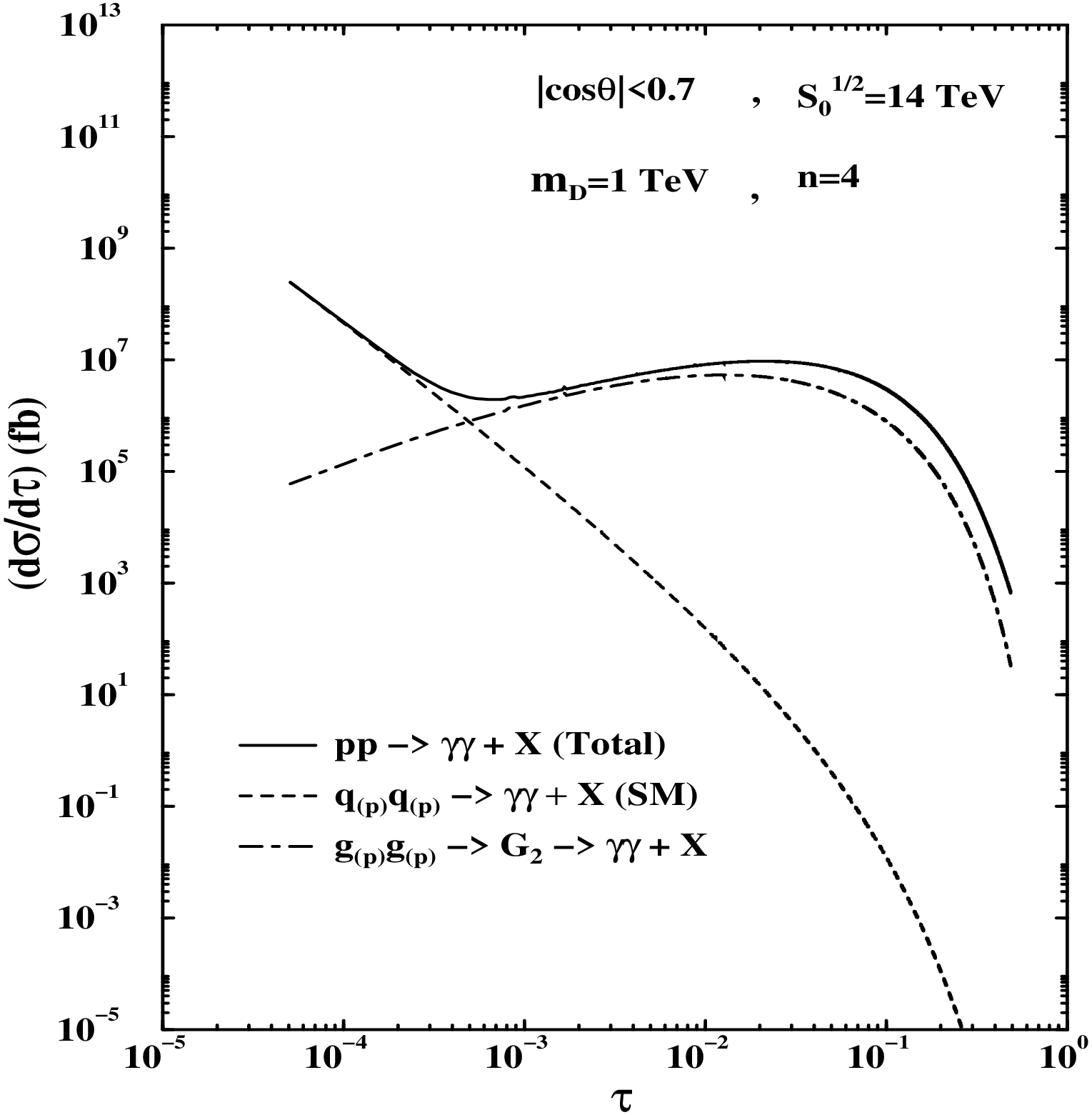,height=8cm,width=8cm,bbllx=0cm,bblly=2cm,bburx=20cm,bbury=25cm,angle=0}
 \end{center}
\caption{\emph{The distribution of $pp\to \gamma\gamma+X$ events as a
function of $\tau$ at the LHC for various sub-processes, where in each case
a cut of $|\cos\theta|<0.7$ is imposed. Shown are
the total (solid line) $pp\to\gamma\gamma +X$ cross section from 
$gg$ and $q \bar q$ fusion processes including the    
graviton exchange and the SM contributions, only the SM 
cross section from $q\bar q\to
\gamma\gamma$ (dashed line) and 
only the graviton exchange cross section 
from $gg\to
\gamma\gamma$ (dot-dashed line). The
model parameters are the same as in Fig.~\ref{fig2}.}}
\label{fig4}
\end{figure}

\begin{figure}[htb]
\psfull
 \begin{center}
 \leavevmode
 \epsfig{file=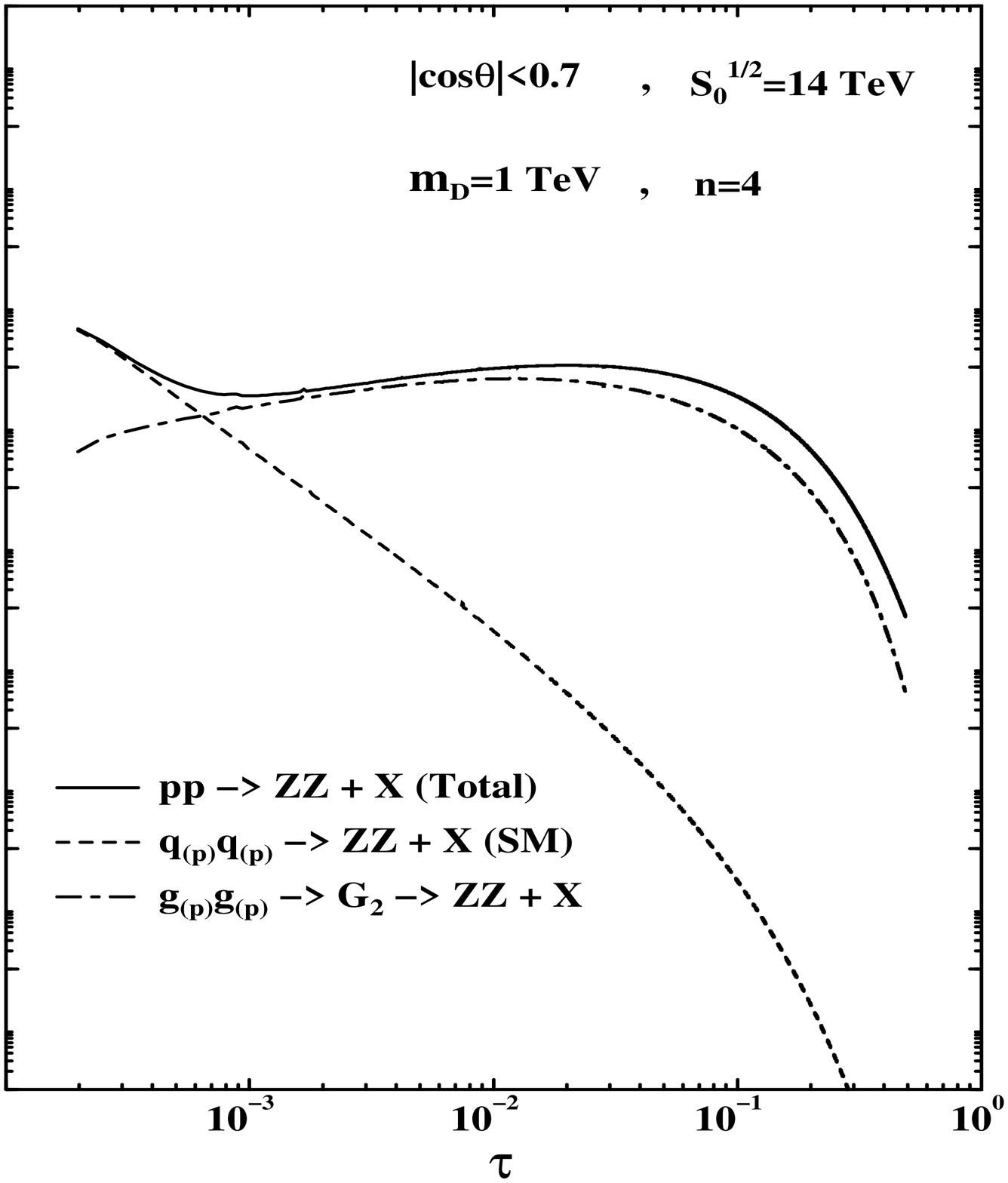,height=8cm,width=8cm,bbllx=0cm,bblly=2cm,bburx=20cm,bbury=25cm,angle=0}
 \end{center}
\caption{\emph{The distribution of $pp\to ZZ+X$ events as a
function of $\tau$ at the LHC for various sub-processes, where in each case
a cut of $|\cos\theta|<0.7$ is imposed. Shown are
the total (solid line) $pp\to ZZ +X$ cross section from 
$gg$ and $q \bar q$ fusion processes including the    
graviton exchange and the SM contributions, only the SM 
cross section from $q\bar q\to ZZ$ (dashed line) and 
only the graviton exchange cross section 
from $gg\to ZZ$ (dot-dashed line). The
model parameters are the same as in Fig.~\ref{fig2}.}}
\label{fig5}
\end{figure}

At hadronic colliders such as the LHC, gauge boson pairs may be produced
via graviton exchange in gluon gluon collisions. In particular, let us 
first consider  $gg\to \gamma \gamma$ and $gg\to ZZ$. 
The parton cross section for $gg\to \gamma\gamma$
is the same as $\gamma\gamma\to gg$ reduced by a factor 
of 64 because of color averaging 
(see appendix B).
Likewise, the parton cross section for
$gg\to ZZ$ is the same as $\gamma\gamma\to ZZ$ reduced by 8 to take into
account color (see appendix D). 
The differential cross section as a function of $\tau$ is shown in 
Fig.~\ref{fig4}
in the case of $\gamma\gamma$ production 
and in Fig.~\ref{fig5} in the case of $ZZ$ production,
where a cut of $|z|<0.7$ has been applied.

In both cases, there is also a contribution from 
$q\bar q\to \gamma\gamma$, $ZZ$ due to $s$-channel graviton exchange. 
These $q \bar q$ annihilation processes also have a SM 
contribution. The 
differential cross sections for 
$q\bar q\to \gamma\gamma$ and $q \bar q \to ZZ$, including the
SM, graviton mediated and SM$\times$graviton interference terms, 
are given in appendix H.      

We note that 
the effects of graviton exchanges at the Tevatron via the 
processes $gg \to \gamma \gamma$ and 
$q\bar q\to \gamma\gamma$ were studied in detail by Cheung in 
\cite{4gammapaper}. Here we focus instead 
on the LHC in which case the dominant 
contribution to $\gamma \gamma +X$ production comes from the gluon fusion 
sub-process as opposed to the Tevatron where $q\bar q\to \gamma\gamma$ 
is more important. As will be shown below, the attainable limit 
on $m_D$ at the LHC from 
$pp \to \gamma \gamma +X$ is much stronger than the one obtained by 
Cheung in \cite{4gammapaper}.
    
In Figs.~\ref{fig4} and \ref{fig5} we plot 
the invariant mass distribution, $d\sigma/d\tau$, for 
the total cross section 
(i.e., including the SM and graviton contributions from 
$gg \to \gamma \gamma,~ZZ$ and 
$q\bar q \to \gamma \gamma,~ZZ$), the SM background from 
$q\bar q \to \gamma \gamma,~ZZ$ and the graviton 
cross section due to the gluon 
fusion process only. Here and in what follows, the corresponding overall  
cross sections of the colliding protons are calculated using  
the CTEQ4M parton distributions \cite{cteq4m}. 
We see that the gluon fusion process is
the dominant production mechanism from graviton exchanges.
Moreover, as in the case of the 
electron-positron collider, the cross section peaks
at relatively large $\tau$ in marked contrast to the SM background which
falls off sharply with $\tau$. Clearly, events with such large $\tau$
would be a distinctive signature of new physics at the LHC with negligible 
SM background.

In Fig~\ref{fig7} we show the $3\sigma$ limits that can be placed 
on the scale of the gravitational interactions $m_D$ at the LHC, by 
measuring $pp \to \gamma \gamma,~ZZ ~+X$.
The limits are obtained by requiring: 

\begin{eqnarray}
\frac{\sigma^T_{M_{VV}^{min}} -\sigma^{SM}_{M_{VV}^{min}}}
{\sqrt{\sigma^T_{M_{VV}^{min}}}} 
\times \sqrt L > 3 ~,\label{bound}
 \end{eqnarray}

\noindent where $\sigma^T_{M_{VV}^{min}}$ is the total cross section
for $\gamma \gamma$ or 
$ZZ$ production at the LHC integrated from a lower $VV$ invariant mass 
cut of $M_{VV}^{min}$ ($V=\gamma$ or $Z$),  
and $\sigma^{SM}_{M_{VV}^{min}}$ is the corresponding 
SM cross section for these processes. Also, we take an integrated luminosity 
of $L=30$ fb$^{-1}$ and we require at least 10 such events above the SM 
background for the given value of $m_D^{min}$ - the lower bound on $m_D$ 
(see \cite{footbound}). 

\begin{figure}[htb]
\psfull
 \begin{center}
 \leavevmode
 \epsfig{file=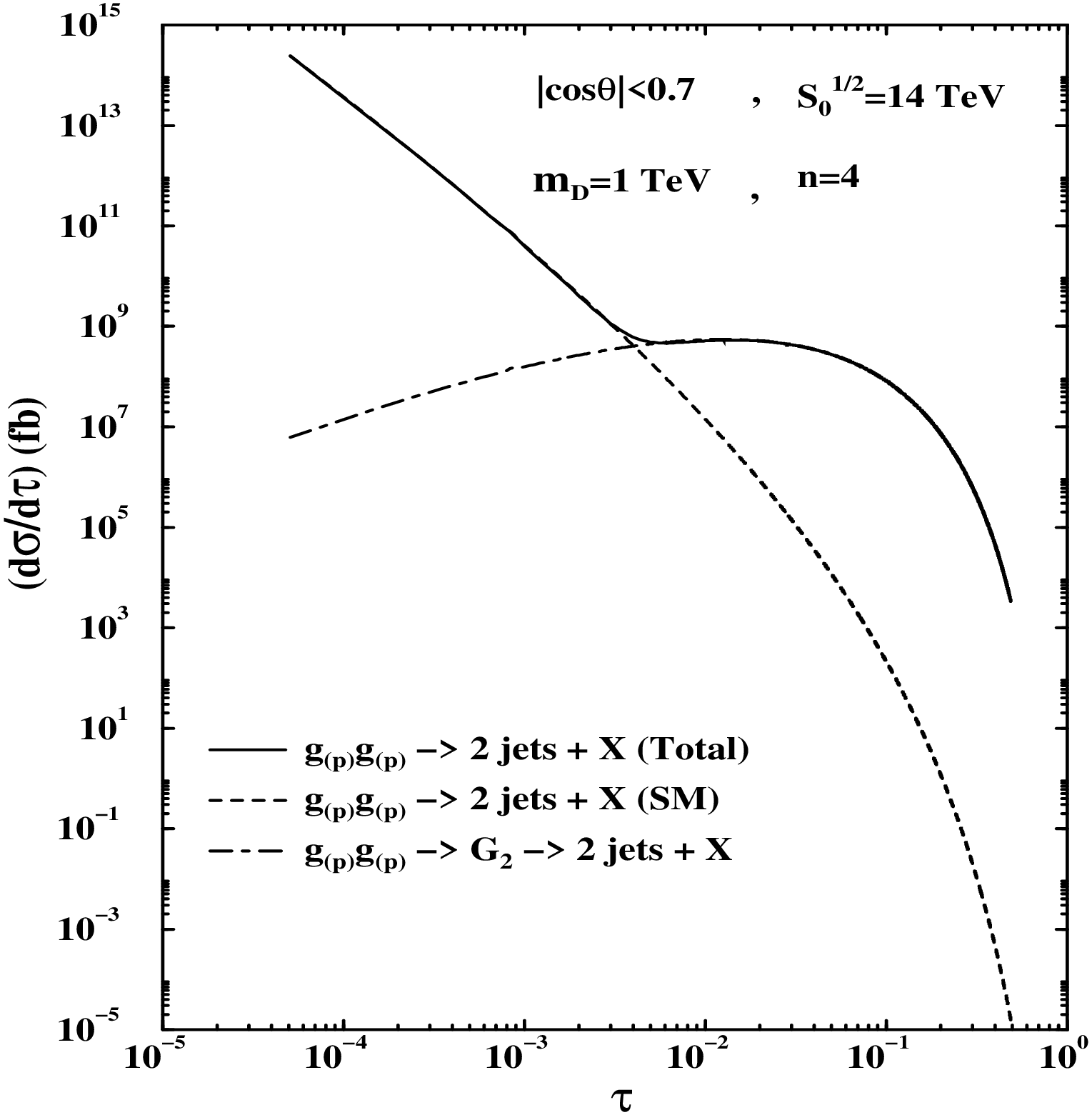,height=8cm,width=8cm,bbllx=0cm,bblly=2cm,bburx=20cm,bbury=25cm,angle=0}
 \end{center}
\caption{\emph{The 2 jets invariant mass 
distribution of $pp\to 2 ~{\rm jets}~+X$ events as a
function of $\tau$ at the LHC for various sub-processes where in each case
a cut of $|\cos\theta|<0.7$ is imposed. Shown are
the total (solid line) $pp\to2~{\rm jets}~+X$ cross section from 
the $gg$ fusion processes $gg \to gg,~q \bar q$ including the    
graviton exchange and the SM contributions, only the SM 
cross section from $gg \to gg,~q \bar q$ (dashed line) and 
only the graviton exchange cross section 
from four gluons scattering sub-process $gg \to gg$ (dot-dashed line). The
model parameters are the same as in Fig.~\ref{fig2}.}}
\label{fig6}
\end{figure}

As stated before, since the SM cross sections for these type of 
processes drop sharply 
with $M_{VV}$, it is advantageous  
to study these cross sections at high $M_{VV}$
in which case one is eliminating a large portion of the SM background. 
Clearly, as can be seen in Fig.~\ref{fig7}, there is an optimal 
lower cut, $M_{VV}^{min}$, that one can impose on these signals in order 
to place the best limits on $m_D$ in case no deviation from the SM 
is observed. 
This is because as one further goes to higher 
$M_{VV}$ values, the gravitational signal sharply falls as well. 
For example, for  $\gamma \gamma$ and 
$ZZ$ production, the optimal lower cuts to be considered are 
$M_{\gamma \gamma}^{min} \sim M_{ZZ}^{min} \approx 2$ TeV, in which case
the obtainable bound at the LHC (with $L=30$ fb$^{-1}$) 
will be $m_D \gsim 6$ TeV. 
We note that, for $m_D=6$ TeV and $M_{\gamma \gamma}^{min} = 2$ TeV, 
$\sigma^T_{M_{\gamma \gamma}^{min}}(pp \to \gamma \gamma+X) \simeq 
0.7$ fb and 
$\sigma^{SM}_{M_{\gamma \gamma}^{min}}(pp \to \gamma \gamma+X) \simeq 
0.1$ fb, yielding about $\sim 20$ $\gamma \gamma$ events out of which 
$\sim 17$ are due to gravitational interactions. Also, 
this limit is more than four times larger than the one found by 
Cheung in \cite{4gammapaper} for the upgraded Tevatron run II case. 
The corresponding cross sections for $ZZ$ production with 
$m_D=6$ TeV and $M_{ZZ}^{min} = 2$ TeV are:
$\sigma^T_{M_{ZZ}^{min}}(pp \to ZZ+X) \simeq 
0.9$ fb 
and 
$\sigma^{SM}_{M_{ZZ}^{min}}(pp \to ZZ+X) \simeq 
0.25$ fb, yielding about $\sim 30$ $ZZ$ events out of which 
$\sim 20$ are due to gravitational interactions.

\begin{figure}[htb]
\psfull
 \begin{center}
 \leavevmode
 \epsfig{file=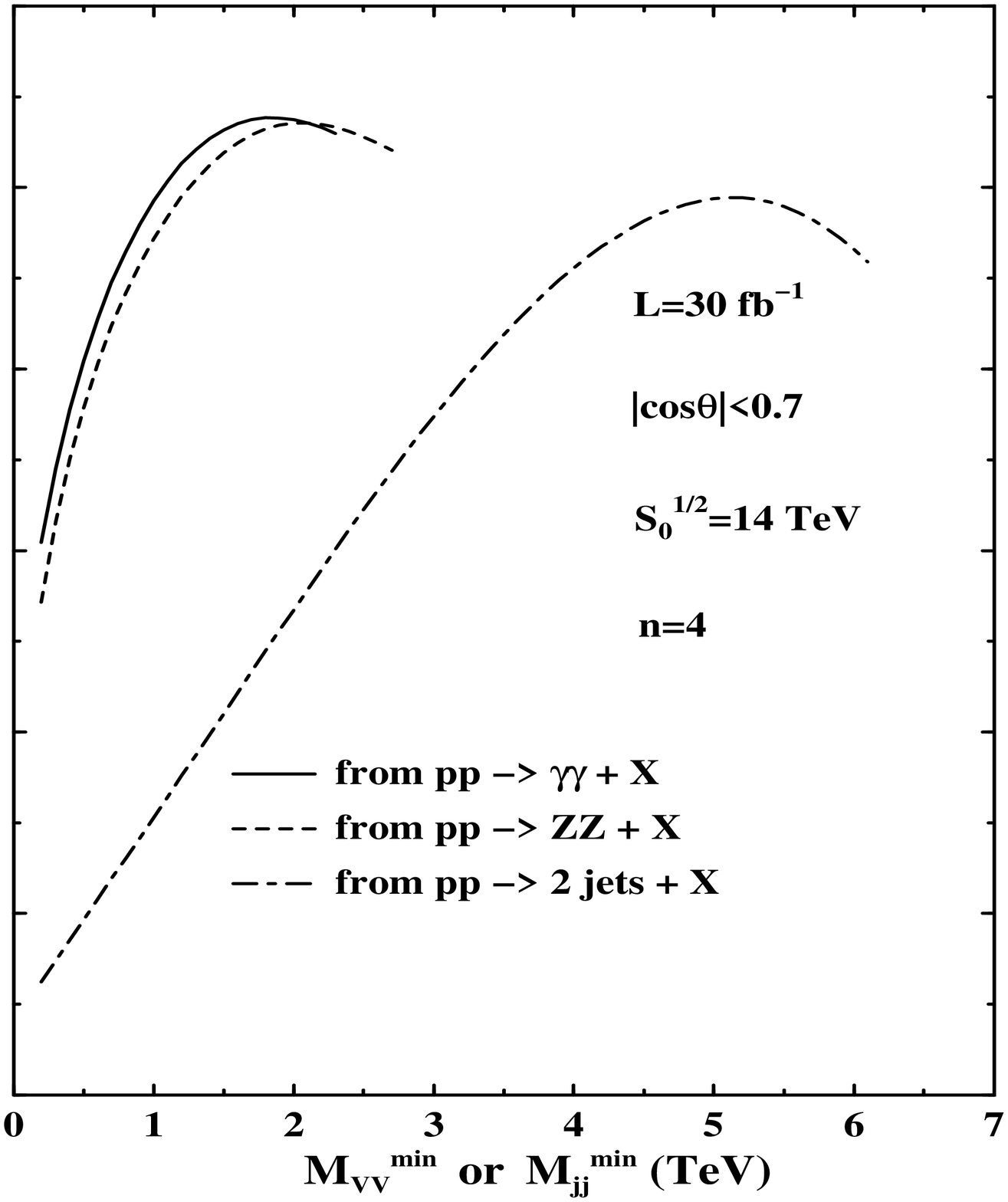,height=8cm,width=8cm,bbllx=0cm,bblly=2cm,bburx=20cm,bbury=25cm,angle=0}
 \end{center}
\caption{\emph{$3\sigma$ bounds on the scale $m_D$ 
of the gravitational interactions derived from eq.~(\ref{bound}), 
as 
a function of the lower cut on the invariant mass $M_{VV}^{min}$ or 
$M_{jj}^{min}$ of the $VV$ or $jj$ system ($V=\gamma$ or $Z$ and $j$=jet), 
see text. Shown are the limits ($m_D^{min}$) that can be obtained by 
using the reactions $pp \to \gamma \gamma +X$ (solid line),  
$pp \to ZZ +X$ (dashed line) and 
$pp \to 2~{\rm jets} +X$ (dot-dashed line), 
where in each case
a cut of $|\cos\theta|<0.7$ is imposed.
The bounds are given for a LHC with an integrated luminosity 
of $30$ inverse fb's. The rest of the 
model parameters are the same as in Fig.~\ref{fig2}.
See also \cite{footbound}.}}
\label{fig7}
\end{figure}

The process $gg\to
W^+W^-$ would be about twice as large as the $gg \to ZZ$ 
due to the Bose symmetry
factor in the latter. The final state however would be more difficult to
observe experimentally.

Another process which may be studied at hadronic colliders is 
$pp~{\rm or}~p \bar p \to 2~{\rm jets}+X$. 
At the LHC, $gg\to gg$ and $gg \to q \bar q$ will be the dominant production 
mechanism of 2 jets due to the large gluon content in the colliding 
protons. These processes can proceed via graviton exchanges where, again, 
the dominant graviton signal comes from the gluon -- gluon scattering 
sub-process $gg\to gg$.  
There will of course be a large 2 jets QCD background 
from the SM. 
The SM and gravity mediated 
amplitudes and cross sections for $gg\to gg$ and $gg \to q \bar q$ 
are given in 
appendix A and H, respectively.

In Fig.~\ref{fig6} we plot the invariant mass distribution, $d\sigma/d\tau$, 
of the total cross section for 
$pp \to 2~{\rm jets}+X$, of the SM QCD background and of  
the graviton cross section due only to the pure four 
gluon scattering process.  
Evidently, the QCD background is again 
peaked at low $\tau$ while the 2 jet events 
resulting from graviton exchange occur at large $\tau$.

In Fig~\ref{fig7} we also show the $3\sigma$ limits (obtained 
from eq.(\ref{bound})) that can be placed 
on $m_D$ at the LHC, by 
measuring the dijets events $pp \to 2~{\rm jets}~+X$.
For reasons mentioned above, there is again an optimal value of 
$M_{jj}^{min}$ ($M_{jj}$ is the invariant mass of the 2 jets system) 
which gives the best limit on $m_D$. 
In particular, in this case, cutting the cross section from below by 
$M_{jj}^{min} \sim 5$ TeV, will result in the $3 \sigma$ 
limit $m_D \gsim 6$ TeV, if no 
deviation from the SM cross section is observed.
Again we note that for $m_D=6$ TeV and $M_{jj}^{min} = 5$ TeV, 
$\sigma^T_{M_{jj}^{min}}(pp \to 2~{\rm jets}+X) \simeq 
1.5$ fb and 
$\sigma^{SM}_{M_{jj}^{min}}(pp \to 2~{\rm jets}+X) \simeq 
0.8$ fb, yielding about $\sim 45$ $jj$ events out of which 
$\sim 20$ are due to gravitational interactions.

\section{Summary and conclusion}

We have studied gauge boson - gauge boson scattering processes 
of the form 
$V_1V_2 \to V_3V_4$, where $V_i$ are SM gauge bosons,  
due to 
graviton exchanges. 
We have shown that these type of processes can lead to some very distinct 
signatures of gravitational interactions at future colliders,
which have no SM background at 
lowest order in the relevant gauge couplings.
 
For example, at an high energy 
electron-positron collider, 
vector bosons which are produced nearly on-shell and 
collinear with the initial particles, can collide and exchange spin 2 and/or 
spin 0 gravitons, leading to appreciably large new signals 
such as $\gamma \gamma \to \gamma \gamma,~
gg,~ZZ$, $\gamma Z \to \gamma Z$, $ZZ\to ZZ$ etc...
The latter is of particular interest since it has some sensitivity to 
scalar graviton excitations if, for some reason, these turn out to be 
enhanced. 

Similarly, gauge boson scattering, due to graviton exchanges, which involve
 two or four gluons, such as 
$gg \to \gamma \gamma,~ZZ,~WW,~gg$, can lead to significantly  
enhanced signals of $\gamma \gamma,~ZZ$, $WW$ and $jj$ 
($j$=jet) pair production at the LHC.

A key feature of all these type of scattering processes is that 
the gravity mediated cross-sections peak at high values of 
the invariant $VV$ mass, whereas their corresponding SM 
cross sections
(which in some instances 
arise only at higher orders in the gauge couplings) are concentrated at 
low $\tau$ values. 
Thus, these type of low energy gravity signals will be quite distinctive. 

Alternatively, if no such new signals are observed, then some of 
 these processes 
can be used to place a bound on the scale of gravitational interactions. 
For example, we find that a limit of $m_D \gsim 3$ TeV can be placed 
on the scale of the low energy gravity using the reaction 
$e^+e^- \to \gamma \gamma e^+e^-$ which proceed predominantly through 
graviton exchanges in the  
$\gamma \gamma \to \gamma \gamma$ sub-process
at the $e^+e^-$ Next Linear Collider. 

In those cases which have a significant SM background, 
we utilized the growth of the gravitational cross-sections 
with $\tau$ to derive the best (optimal) limits. 
For example, we found that $m_D \gsim 6$ TeV will be obtainable at 
the LHC by measuring the production rates of 
$pp \to \gamma \gamma,~ZZ,~WW~+X$ and $pp \to 2~{\rm jets}~+X$ which, 
at high $\tau$, are driven predominantly by 
graviton exchanges in the
$gg \to \gamma \gamma,~ZZ,~WW,~gg$ sub-processes.

\bigskip
\bigskip

S.B. thanks Jose Wudka for discussions.
This research was supported in part by US DOE Contract Nos. 
DE-FG02-94ER40817 (ISU), DE-FG03-94ER40837 (UCR) and DE-AC02-98CH10886
(BNL).

\newpage
\begin{center}
\noindent {\Large \bf Appendices: Helicity amplitudes and cross sections}
\end{center}

For each of the processes we discuss, we give the helicity amplitudes, 
the differential cross section and the cross section. In all cases the 
scattering is of the general form $V_1V_2\to V_3V_4$ where $V_i$ is a 
vector boson with momentum $p_i$ and helicity $h_i$. The angle between
$\vec p_1$ and $\vec p_3$ in the cms frame is $\theta$ and $z=\cos\theta$.
We also define the quantity $s=(p_1+p_2)^2$.\\

\noindent{\large \bf A: $\gamma\gamma\to\gamma\gamma$ and $gg\to gg$}
\setcounter{num}{1}
\setcounter{equation}{0}
\def\theequation{\Alph{num}.\arabic{equation}}\\

The helicity amplitudes for these processes, in which the graviton exchange 
is in all three $s$, $t$ and $u$-channels, are:

\begin{eqnarray}
&&\fm_{4\gamma}(h1,h2,h3,h4)
=
{1\over 4}\kappa^2 D \times \nonumber \\
&&~~~~~~~~~~\left \{
\begin{array}{ll}
2Q_s  s^2 &{\rm if}~h1=h2=h3=h4  \\
Q_u  (1+z)^2s^2/2 &{\rm if}~h1=-h2=-h3=h4  \\
Q_t  (1-z)^2s^2/2 &{\rm if}~h1=-h2=h3=-h4  \\
0 & {\rm otherwise}
\end{array}
\right .
~, \label{a1}
\end{eqnarray}

\noindent
where 
for $\gamma\gamma\to\gamma\gamma$, $Q_s=Q_t=Q_u=1$ while for 
$gg\to gg$, 
$Q_s=(\delta_{AC}\delta_{BD}+\delta_{AD}\delta_{BC})/2$,
$Q_t=(\delta_{AB}\delta_{CD}+\delta_{AD}\delta_{BC})/2$ and
$Q_u=(\delta_{AC}\delta_{BD}+\delta_{AB}\delta_{CD})/2$,
where $A$, $B$, $C$ and $D$ are the color indices for 
$g_1$
$g_2$
$g_3$ and
$g_4$ respectively.

This leads to the following differential and 
total cross sections, previously derived in~\cite{4gammapaper}:

\begin{eqnarray}
{d\sigma_{2\gamma\to 2\gamma}\over dz}
&=&
{\pi F^2 Q_{tot}\over 16 s}\left ({s^4\over m_D^8}\right ) (z^2+3)^2
~, \label{a2} \\
\sigma_{2\gamma\to 2\gamma}&=&
{\pi F^2 Q_{tot}\over s}
\left (
{s^4\over m_D^8}
\right )
\left ( {7\over 5}\right )
~, \label{a3}
\end{eqnarray}

\noindent
where $Q_{tot}=1$ for $\gamma\gamma\to\gamma\gamma$ and $Q_{tot}=9/16$ for 
$gg\to gg$.

In the case of $gg\to gg$ 
there is also a SM contribution which is given for example in \cite{collider}.
The graviton amplitude, therefore, interferes with 
the tree level SM diagrams. Denoting the interference term by
$\sigma^I_{gg\to gg}$, this interference is given by:

\begin{eqnarray}
{d\sigma^I_{gg\to gg}\over dz}=-
{5\over 2}
\sqrt{{\pi \alpha_s^2\over s} \left ({d \sigma_{gg\to gg}\over d z 
}\right )  } ~~,~~
\sigma^I_{gg\to gg}=-
{25\over 42}\sqrt{35{\pi \alpha_s^2\over s} \sigma_{gg\to gg}   }
~,\label{a4}
\end{eqnarray}

\noindent where $\alpha_s=g_s^2/4 \pi$ and $g_s$ is the QCD coupling.\\

\noindent{\large \bf B: $\gamma\gamma\to gg$}
\setcounter{num}{2}
\setcounter{equation}{0}
\def\theequation{\Alph{num}.\arabic{equation}}\\

In the case of $\gamma\gamma\to gg$ the graviton exchange is only in the 
$s$-channel and the helicity amplitudes are:

\begin{eqnarray}
&&\fm_{2\gamma\to 2g}(h1,h2,h3,h4)
=
{1\over 16}\kappa^2 D \times \nonumber\\
&&~~~~~~~~~~\left \{
\begin{array}{ll}
(1+z)^2s^2\delta_{AB} &{\rm if}~h1=-h2=h4=-h3  \\
(1-z)^2s^2\delta_{AB} &{\rm if}~h1=-h2=h3=-h4  \\
0 & {\rm otherwise}
\end{array}
\right .
~, \label{b1}
\end{eqnarray}

\noindent where $A$ and $B$ are color indices. 

This leads to the differential and total cross sections:

\begin{eqnarray}
{d\sigma_{2\gamma\to 2 g} \over dz}&=&{\pi F^2\over 8 s}\left ( {s^4\over 
m_D^8}\right ) (z^4+6z^2+1)~,\label{b2}\\
\sigma_{2\gamma\to 2 g} &=&
{4\over 5}
{\pi F^2\over s}\left 
( {s^4\over m_D^8}\right )
~.\label{b3}
\end{eqnarray}

\noindent
The cross sections in the reverse reaction $gg\to \gamma\gamma$ are 
smaller by a factor of $64$ due to color averaging 
and was also derived by Cheung in \cite{4gammapaper} with whom we agree. \\

\noindent{\large \bf C: $g\gamma\to g\gamma$}
\setcounter{num}{3}
\setcounter{equation}{0}
\def\theequation{\Alph{num}.\arabic{equation}}\\

In this case, the graviton exchange is only in the $t$-channel and the 
helicity amplitudes are:

\begin{eqnarray}
&& \fm_{g\gamma\to g\gamma}(h1,h2,h3,h4) 
=
{1\over 4}\kappa^2 D \times \nonumber\\
&&~~~~~~~~~~\left \{
\begin{array}{ll}
s^2 &{\rm if}~h1=h2=h3=h4  \\
(1+z)^2 s^2/4 &{\rm if}~h1=-h2=h3=-h4  \\
0 & {\rm otherwise}
\end{array}
\right .
~, \label{c1}
\end{eqnarray}

\noindent which gives a differential and total cross section of:

\begin{eqnarray}
{d\sigma_{g\gamma\to g\gamma}\over dz}
&=&
{\pi F^2\over 64 s}\left ({s^4\over m_D^8}\right ) (17+4z+6z^2+4z^3+z^4)
~,\label{c2}\\
\sigma_{g\gamma\to g\gamma}&=&
{\pi F^2\over s} 
\left ({s^4\over m_D^8}\right )
\left ( {3\over 5}\right )
~.\label{c3}
\end{eqnarray}\\

\noindent{\large \bf D: $\gamma\gamma\to ZZ$ and $gg\to ZZ$}
\setcounter{num}{4}
\setcounter{equation}{0}
\def\theequation{\Alph{num}.\arabic{equation}}\\

In the case of $\gamma\gamma\to ZZ$ the graviton exchange is also only 
in the $s$-channel and 
the helicity amplitudes are given by: 

\begin{eqnarray}
\fm_{2\gamma\to 2Z}(h1,h2,h3,h4)
=
{1\over 4}\kappa^2 D s^2 \alpha_{h1,h2,h3,h4}(z,x_Z)
~,\label{d1}
\end{eqnarray}

\noindent
where $x_Z=m_Z^2/s$ and we have defined the ``reduced'' 
helicity amplitude $\alpha_{h1,h2,h3,h4}$ which is given in Table 1.
Note that the amplitude is only non-zero if $h1=-h2$ so only 
$\alpha_{+1,-1,h3,h4}$ is given. The case where $h1=-1$ is clearly 
related by Parity.\\

$$
\begin{tabular}{|c|c|}
\hline
$h1~h2~h3~h4$ & $\alpha_{h1,h2,h3,h4}(z,x_Z)$ \\
\hline
\hline
$+-\pm\pm$,   
& $x_Z(1-z^2)$\\
\hline
$+-\pm 0$;
$+-0\mp$
& $(x_Z(1-z^2)/2)^{1/2}(1\pm z)$\\
\hline
$+-\pm\mp$
& $(1\pm z)^2/4$\\
\hline
$+-00$
& $-(1-z^2)(1+4x_Z)/4$ \\
\hline
\end{tabular}
$$

\bigskip
\bigskip

{\bf Table 1:} {\emph {The reduced helicity amplitude 
$\alpha_{h1,h2,h3,h4}$, as defined 
in eq.(\ref{d1}), 
for the process $\gamma \gamma \to ZZ$ or $gg \to ZZ$. 
Values not given are either zero or related to the given values by Parity.}}

\bigskip
\bigskip

The differential and total cross sections for this process are thus given by:

\begin{eqnarray}
{d\sigma_{2\gamma\to 2Z}\over dz}
&=&
{\pi F^2\over 128 s}
\left ({s^4\over m_D^8}\right ) 
\beta_Z (3+4x_Z-4x_Zz^2+z^2)
\nonumber\\
&&
(1+12x_Z-12z^2x_Z+3z^2)
\label{d2}\\
\sigma_{2\gamma\to 2Z}&=&
{\pi F^2\over 120 s}  
\left ({s^4\over m_D^8}\right )
\beta_Z (48 x_Z^2 +56 x_Z+13)
~,\label{d3}
\end{eqnarray}

\noindent where:

\begin{eqnarray}
\beta_Z=\sqrt{1-4x_Z} \label{betaz} ~.
\end{eqnarray}
 
\noindent 
In the case of $gg\to ZZ$ the cross sections are related to the above by 
$\sigma_{2g\to 2Z}=\sigma_{2\gamma\to 2Z}/8$ due to color. \\

\noindent{\large \bf E: $ZZ\to\gamma\gamma$}
\setcounter{num}{5}
\setcounter{equation}{0}
\def\theequation{\Alph{num}.\arabic{equation}}\\

The process $ZZ\to\gamma\gamma$
is the reverse of $\gamma\gamma\to ZZ$ 
so the matrix elements are the same as the reverse process, however 
in order to analyze the cross section using the effective boson 
approximation we need to obtain the differential and 
total cross section for each initial 
helicity state. These are given by:

\begin{eqnarray}
&&\frac{d\sigma_{2Z\to 2\gamma}}{d z} (h_1,h_2) = 
{\pi F^2\over 4 s}
\left ({s^4\over m_D^8}\right ) \beta_Z^{-1} \times \nonumber\\ 
&&~~~~~~~~\left \{
\begin{array}{ll}
(1+6z^2+z^4)/8 & {\rm for }~h1,h2=\pm,\mp \\
2x_Z^2(1-z^2)^2 & {\rm for }~h1,h2=\pm,\pm \\
x_Z(1-z^4) & {\rm for }~h1,h2=\pm, 0;~0,\pm\\
(1+4x_Z)^2 (1-z^2)^2/8 & {\rm for }~h1,h2=0,0
\end{array}
\right.
~,\label{e1}
\end{eqnarray}

\begin{eqnarray}
&&\sigma_{2Z\to 2\gamma}(h_1,h_2) = 
{\pi F^2\over 4 s}
\left ({s^4\over m_D^8}\right ) \beta_Z^{-1} \times \nonumber\\
&&~~~~~~~~\left \{
\begin{array}{ll}
4/5 & {\rm for }~h1,h2=\pm,\mp \\
32 x_Z^2/15 & {\rm for }~h1,h2=\pm,\pm \\
8 x_Z/5     & {\rm for }~h1,h2=\pm, 0;~0,\pm\\
2(1+4x_Z)^2/15 & {\rm for }~h1,h2=0,0
\end{array}
\right.
~, \label{e2}
\end{eqnarray}

\noindent where $\beta_Z$ is defined in eq.~(\ref{betaz}) and 
again $x_Z=m_Z^2/s$.\\

\noindent{\large \bf F: $\gamma Z \to \gamma Z$}
\setcounter{num}{6}
\setcounter{equation}{0}
\def\theequation{\Alph{num}.\arabic{equation}}\\

In the case of $\gamma Z\to \gamma Z$ the reaction proceeds only through 
the $t$-channel. The helicity amplitudes for this process are:

\begin{eqnarray}
\fm_{\gamma Z \to \gamma Z}(h1,h2,h3,h4)
=
{1\over 4}\kappa^2 D s^2 \beta_{h1,h2,h3,h4}(z,x_Z)
~,\label{f1}
\end{eqnarray}

\noindent
where the reduced helicity amplitudes in this case, $\beta_{h1,h2,h3,h4}$, 
are given 
in Table 2.\\

$$
\begin{tabular}{|c|c|}
\hline
$h1~h2~h3~h4$ & $\beta_{h1,h2,h3,h4}(z,x_Z)$ \\
\hline
\hline
$++++$   
&$ (2-x_Z+x_Zz)^2(1-x_Z)^2/4 $\\
\hline
$+-++$   
&$ x_Z(1-x_Z)^2(1-z^2)/4 $\\
\hline
$+-+-$   
&$ (1-x_Z)^2(1+z)^2/4 $\\
\hline
$+++0$   
&$ -\sqrt{2x_Z(1-z^2)}(1-x_Z)^2(2-x_Z+zx_Z)/4 $\\
\hline
$+0+-$   
&$ -\sqrt{2x_Z(1-z^2)}(1-x_Z)^2(1+z)/4   $\\
\hline
$+0+0$   
&$ (1+z)(1-x_Z)^2(1-x+x_Zz)/2 $\\
\hline
\end{tabular}
$$

\bigskip
\bigskip

{\bf Table 2:} {\emph {The reduced helicity amplitude 
$\beta_{h1,h2,h3,h4}$, as defined 
in eq.(\ref{f1}),
for the process $\gamma Z \to \gamma Z$.
Values not given are either zero or related to the given 
values by Parity.}}

\bigskip
\bigskip

The total cross section as a function of the initial $Z$ helicity 
and averaged over the initial photon polarizations is: 

\begin{eqnarray}
&&\sigma_{\gamma Z\to \gamma Z}(h2)=
{\pi F^2\over 60 s} \left ( {s^4\over m_D^8} \right )
\times \nonumber\\
&&\left \{
\begin{array}{cl}
(1-x_Z)^4(36-47x_Z+52x_Z^2-27x_Z^3+6x_Z^4) 
&~~~{\rm for}~~h2=\pm\\ 
2 (1-x_Z)^4(10+3x_Z-6x_Z^2+3x_Z^3)
&~~~{\rm for}~~h2=0 
\end{array}
\right .
~, \label{f2} \nonumber\\
~
\end{eqnarray}\\

\noindent{\large \bf G: $ZZ \to ZZ$}
\setcounter{num}{7}
\setcounter{equation}{0}
\def\theequation{\Alph{num}.\arabic{equation}}\\

This process proceeds in all three channels via both spin 2 and spin 0 
graviton exchanges and, in the SM, via neutral Higgs exchange. 
For the graviton exchange the helicity amplitudes for this process are:

\begin{eqnarray}
\fm_{2Z \to 2Z}(h1,h2,h3,h4)
=
{1\over 4}\kappa^2 D s^2 \gamma_{h1,h2,h3,h4}(z,x_Z,R)
~,\label{g1}
\end{eqnarray}

\noindent
where the reduced helicity amplitudes $\gamma_{h1,h2,h3,h4}$ are 
given in Table~3 and 
$R=2(1+\epsilon)(n-1)/(3n+6)$ 
is the factor associated with the scalar 
propagator as discussed in the text.

$$
\begin{tabular}{|c|c|c|}
\hline
$h1~h2~h3~h4$ & $\gamma_{h1,h2,h3,h4}^2(z,x_Z)$ & $\gamma_{h1,h2,h3,h4}^0(z,x_Z)$  \\
\hline
\hline
$+-\pm\mp$
& $(3-6x_Z+8 x_Z^2)(1\pm z)^2/6$
& $2 x_Z^2 (1\pm z)^2$
\\
\hline
$+-\pm\pm$
& $x_Z(9-4x_Z)(1-z^2)/6$
& $x_Z^2(1-z^2)$
\\
\hline
$\pm\pm\pm\pm$ 
& $2(24x_Z^2+8x_Z^2 z^2-18x_Z+3)/3$
& $2x_Z^2(z^2+3)$
\\
\hline
$\pm\pm\mp\mp$
& $16x_Z^2z^2/3 $
& $2x_Z^2(3+z^2)$
\\
\hline
$0000$
& $ (3+6z^2x_Z-22x_Z+24x_Z^2z^2+24x_Z^2+z^2)/3$
& $(24x_Z^2-16 x_Z+3+z^2)$
\\
\hline
$00 \pm\pm$
& $x_Z(15-28x_Z-7z^2+36z^2x_Z)/6$
& $x_Z(3-4x_Z-z^2)$
\\
\hline
$00 \pm\mp$
& $-(1-z^2)(3+26 x_Z-24 x_Z^2)/12$
& $-x_Z(1-z^2)$
\\
\hline
$000\pm$
& $\pm (7+36x_Z)z\sqrt{2x_Z(1-z^2)}/12$
& $\pm z\sqrt{2x_Z(1-z^2)}/2$
\\
\hline
$0+0\pm;~0+\mp 0$
& $(1\pm z)(3-28 x_Z+32x_Z^2\pm 16x_Zz)/6$
& $-x_Z(1\pm z)(1-2x_Z\mp z)$
\\
\hline
$0+\pm\pm$
& $\pm 8 z\sqrt{2x_Z^3(1-z^2)}$
& $\pm   z\sqrt{2x_Z^3(1-z^2)}$
\\
\hline
$0+\mp\pm$
& $-(1\pm z)(9-4x_Z)\sqrt{2x_Z(1-z^2)}/12$
& $-(1\pm z)x_Z \sqrt{2x_Z(1-z^2)} $
\\
\hline
\end{tabular}
$$

\bigskip
\bigskip

{\bf Table 3:} {\emph {The reduced helicity amplitude 
$\gamma_{h1,h2,h3,h4}$, as defined 
in eq.(\ref{g1}),  
for the process $ZZ \to ZZ$. 
We have defined 
$\gamma_{h1,h2,h3,h4}(z,x_Z,R)=\gamma_{h1,h2,h3,h4}^2(z,x_Z)+
R\gamma_{h1,h2,h3,h4}^0(z,x_Z)$, where 
$R=2(1+\epsilon)(n-1)/(3n+6)$. See also text. Values 
not given are related to the given 
values by Parity.}}

\bigskip
\bigskip

For the SM Higgs boson exchange the helicity amplitudes for this process are: 

\begin{eqnarray}
\fm_{2Z \to 2Z}^{SM}(h1,h2,h3,h4)
=
{1\over 8} g_W^2 \frac{m_Z^2}{m_W^2} s \gamma_{h1,h2,h3,h4}^{SM}(z,x_Z,\Pi_s,\Pi_t,\Pi_u)
~,\label{g1sm}
\end{eqnarray}

\noindent where $g_W=e/s_W$ is the weak coupling and 
$\Pi_x=(x - m_H^2 +i \Gamma_H m_H)^{-1}$, $x=s,~t,~u$,  
are the $s$, $t$ and $u$-channel factors associated with 
the corresponding SM Higgs propagators, where 
$t(u)=s(z -(+) 1)(4x_Z-1)/2$.   
The reduced SM helicity amplitudes $\gamma_{h1,h2,h3,h4}^{SM}$ are 
given in Table~4.

$$
\begin{tabular}{|c|c|}
\hline
$h1~h2~h3~h4$ & $\gamma_{h1,h2,h3,h4}^{SM}(z,x_Z,\Pi_s,\Pi_t,\Pi_u)$ \\
\hline
\hline
$+-\pm\mp$   
&$ 2 x_Z(1 \pm z)^2 (\Pi_t+\Pi_u)$\\
\hline
$+-\pm \pm$   
&$ 2 x_Z(1 - z^2) (\Pi_t+\Pi_u) $\\
\hline
$++ \pm \pm$   
&$ 2x_Z \left[ 4\Pi_s + (1 \pm z)^2 \Pi_t + (1 \mp z)^2 \Pi_u \right]$\\
\hline
$0000$   
&$ 2 \left[ 4\Pi_s (1-2x_Z)^2 +  \Pi_t (z-1+4x_Z)^2 + \Pi_u (z+1-4x_Z)^2 \right]/x_Z$\\
\hline
$00\pm \pm$   
&$  \left[ 4\Pi_s (1-2x_Z) +  (1-z^2)(\Pi_t+\Pi_u) \right]   $\\
\hline
$00 \pm \mp $   
&$ -(1-z^2)(\Pi_t+\Pi_u) $\\
\hline
$00 0\pm $   
&$ \pm \sqrt{2x_Z(1-z^2)} \left[ \Pi_t(z-1+4x_Z) + \Pi_u(z+1-4x_Z)\right]/2x_Z $\\
\hline
$0 + \pm 0 $   
&$ -(1 \mp z) \left[ \Pi_t(z \pm 1) + \Pi_u(z+1-4x_Z)\right]$\\
\hline
$0 + 0 \pm $   
&$ (1 \pm z) \left[ \Pi_u(z \mp 1) + \Pi_t(z-1+4x_Z)\right]$\\
\hline
$0 + \pm \pm $   
&$ \sqrt{2x_Z(1-z^2)} \left[ \Pi_t(z \pm 1) + \Pi_u(z \mp 1)\right] $\\
\hline
$0 + \mp \pm $   
&$ -\sqrt{2x_Z(1-z^2)} (z \pm 1) (\Pi_t +\Pi_u)/2$\\
\hline
\end{tabular}
$$

\bigskip  
\bigskip

{\bf Table 4:} {\emph {The reduced helicity amplitude 
$\gamma_{h1,h2,h3,h4}^{SM}$ for the SM Higgs exchange contribution 
to the process $Z Z \to Z Z$,
as defined 
in eq.(\ref{g1sm}).
Values not given are related to the given 
values by Parity. $\Pi_{s,t,u}$ are defined in the text above.}}

\bigskip
\bigskip

\noindent{\large \bf H: $q \bar q \to \gamma \gamma$, 
$q \bar q \to ZZ$ and $gg\to q\bar q$}
\setcounter{num}{8}
\setcounter{equation}{0}
\def\theequation{\Alph{num}.\arabic{equation}}\\

Though not gauge-gauge scattering processes, we require 
$q \bar q \to \gamma \gamma$, $q \bar q \to ZZ$ and $gg\to q\bar q$ for the 
processes $p p \to \gamma \gamma +X$, $p p \to ZZ +X$ and $pp \to 2~jets$, 
respectively. 
As in the case of $gg\to gg$, these processes have a SM contribution and, 
therefore, also an 
interference term of the graviton exchange with the SM 
diagrams.\footnote{The differential cross section for 
$q\bar q \to \gamma \gamma$ including the SM and graviton exchanges 
was also derived by Cheung in \cite{4gammapaper}}.

The pure SM differential cross sections for these processes are:

\begin{eqnarray}
{d \sigma_{q\bar q \to \gamma \gamma}^{SM} \over d z}
&=&
\frac{e^4 Q_q^4}{48 \pi s} \left( \frac{1+z^2}{1-z^2} \right)
~,\label{h1} \\
{d \sigma_{q\bar q \to ZZ}^{SM} \over d z}
&=&
\frac{e^4}{24 \pi s}~ \frac{(g_L^q)^4+(g_R^q)^4}{s_W^4(1-s_W^2)^2}~\beta_Z \times \nonumber \\
&&~\frac{2+\beta_Z^2(3-\beta_Z^4)-z^2 \beta_Z^2(9-10 \beta_Z^2+\beta_Z^4)-
4z^4\beta_Z^4}{\left[ (1+\beta_Z^2)^2-4z^2\beta_Z^2 \right]^2}
~,\label{h2}\\
{d \sigma_{g g\to q\bar q}^{SM} \over d z}
&=&
\frac{g_s^4} {1536 \pi s} \left( \frac{7+16 z^2+9z^4}{1-z^2} \right)
~.\label{h3}
\end{eqnarray}
  
\noindent where $\beta_Z$ is defined in eq.~(\ref{betaz}) and 
$s_W=\sin\theta_W$, where 
$\theta_W$ is the weak mixing angle. Also, $g_L^q$ and $g_R^q$ are left and 
right handed couplings of a $Z$-boson to quarks; for $q=u$ (up quark)  
 $g_L^u=1/2-2s_W^2/3$, $g_R^u=-2s_W^2/3$ and 
for $q=d$ (down quark) $g_L^d=-1/2+s_W^2/3$, $g_R^d=-s_W^2/3$.

The pure gravity mediated differential cross sections 
for these processes are:  

\begin{eqnarray}
{d \sigma_{q \bar q \to \gamma \gamma}^G \over d z}
&=&
{\pi F^2\over 192 s} \left ( {s^4\over m_D^8} \right ) (1-z^4)
~,\label{h4}\\
{d \sigma_{q \bar q \to ZZ}^G \over d z}
&=&
{\pi F^2\over 384 s} \left ( {s^4\over m_D^8} \right ) \beta_Z 
\times \nonumber\\
&&~~~~\left[ 4 +3\beta_Z^4 z^2(1-z^2) - 2 \beta_Z^2(1+z^2) \right]
~,\label{h5}\\
{d \sigma_{gg\to q\bar q}^G \over d z}
&=&
\frac{9}{4}{d \sigma_{q \bar q \to \gamma \gamma}^G \over d z}
~,\label{h6}
\end{eqnarray}

\noindent and the corresponding interference terms are:

\begin{eqnarray}
{d \sigma_{q \bar q \to \gamma \gamma}^I \over d z}
&=&
{e^2 Q_q^2 F\over 48 s} \left ( {s^2\over m_D^4} \right ) (1+z^2)
~,\label{h7}\\
{d \sigma_{q \bar q \to ZZ}^I \over d z}
&=&
\frac{e^2 F}{48 s}~ \frac{(g_L^q)^2+(g_R^q)^2}{s_W^2(1-s_W^2)} 
~ \beta_Z \times \nonumber \\
&&~~~~\frac{-2-\beta_Z^2(1-\beta_Z^2)+5 z^2 \beta_Z^2 (1-\beta_Z^2)
+2z^4\beta_Z^4}{(1+\beta_Z^2)^2-4z^2\beta_Z^2}
~,\label{h8}\\
{d \sigma_{gg\to q\bar q}^I \over d z}
&=&
\frac{3}{8}{d \sigma_{q \bar q \to \gamma \gamma}^I \over d z}
~.\label{h9}
\end{eqnarray}

%
%


\begin{thebibliography}{99}
\bibitem{footbound} Notice that the lines in Fig.~\ref{fig7} 
end rather abraptly. This signifies that our 10 events criterion 
can not be met beyond the largest
values of $M_{VV}^{min}$ and/or $M_{jj}^{min}$ shown. 
In these ranges of $M_{VV}^{min}$ and/or 
$M_{jj}^{min}$ we assume that no bound can be attained with a luminosity 
of $L=30$ fb$^{-1}$.
\end{thebibliography}
\end{document}